\pdfoutput=1
\documentclass[11pt,twoside,a4paper,cmspaper,final,collab]{cms-tdr}
\def\svnVersion{b86b6b2}\def\svnDate{2025/08/14}\def\cmsCernNoTag{CERN-EP-2024-332}\def\cmsCernDate{\today}\def\cmsMessage{Published in Physical Review Letters as \href{http://dx.doi.org/10.1103/t5kp-vsv7}{\doi{10.1103/t5kp-vsv7}.}}
\begin{document}\cmsNoteHeader{HIN-23-002}

\newcommand{\pp}    {\ensuremath{\Pp\Pp}\xspace}
\newcommand{\pPb}   {\ensuremath{\Pp\mathrm{Pb}}\xspace}
\newcommand{\PbPb}  {\ensuremath{\text{PbPb}}\xspace}
\newcommand{\noff}  {\ensuremath{N_\text{trk}^\text{offline}}\xspace}
\newcommand{\ncorr}  {\ensuremath{N_\text{trk}^\text{corrected}}\xspace}
\newcommand{\Vz}    {\ensuremath{\mathrm{V}^\text{0}}\xspace}
\newcommand{\vsub}  {\ensuremath{v_2\{4,\text{Sub}\}}\xspace}
\newcommand{\llangle}{\ensuremath{\langle\langle}}
\newcommand{\rrangle}{\ensuremath{\rangle\rangle}}
\newcommand{\ptpoi}  {\ensuremath{p_\text{T}^\text{POI}}\xspace}
\newcommand{\ngen}  {\ensuremath{N_\text{trk}^\text{gen}}\xspace}
\providecommand{\cmsTable}[1]{\resizebox{\textwidth}{!}{#1}}
\ifthenelse{\boolean{cms@external}}{\providecommand{\cmsLeft}{upper\xspace}}{\providecommand{\cmsLeft}{left\xspace}}
\ifthenelse{\boolean{cms@external}}{\providecommand{\cmsRight}{lower\xspace}}{\providecommand{\cmsRight}{right\xspace}}
\ifthenelse{\boolean{cms@external}}{\providecommand{\NA}{\ensuremath{\cdots}\xspace}}{\providecommand{\NA}{\ensuremath{\text{---}}\xspace}}
\def\appendixname{Back matter}
\newcommand{\appname}{\appendixname}

\cmsNoteHeader{HIN-23-002}
\title{Evidence for similar collectivity of high transverse momentum particles in \texorpdfstring{\pPb}{pPb} and \texorpdfstring{\PbPb}{PbPb} collisions}

\author*[cern]{A. Cern Person}

\date{\today}

\abstract{
Charged hadron elliptic anisotropies ($v_2$) are presented over a wide transverse momentum (\pt) range for proton-lead (\pPb) and lead-lead (\PbPb) collisions at nucleon-nucleon center-of-mass energies of 8.16 and 5.02\TeV, respectively. The data were recorded by the CMS experiment and correspond to integrated luminosities of 186\nbinv and 0.607\nbinv for the \pPb and \PbPb systems, respectively. A four-particle cumulant analysis is performed using subevents separated in pseudorapidity to effectively suppress non-collective effects. 
At high \pt ($\pt>8\GeV$), significant positive $v_2$ values are observed that are similar between \pPb and \PbPb collisions at comparable charged particle multiplicities. This observation suggests a common origin for the multi-particle collectivity for high-\pt particles in the two systems.}

\hypersetup{%
pdfauthor={CMS Collaboration},%
pdftitle={Elliptic anisotropy at high pT in pPb collisions using subevent cumulants},%
pdfsubject={CMS},%
pdfkeywords={CMS, heavy ion, high pT flow, small systems, cumulant}}

\maketitle 

Strong azimuthal correlations among particles emitted over a wide range of polar angles in relativistic nucleus-nucleus (AA) collisions at the BNL RHIC~\cite{Adams:2005ph,Abelev:2009af,Alver:2009id} and 
CERN LHC~\cite{Chatrchyan:2011eka,Chatrchyan:2012wg,Aamodt:2010pa,ATLAS:2012at,Chatrchyan:2012ta,CMS:2013bza,CMS:2024krd} colliders indicate the formation of a quark-gluon plasma (QGP) that exhibits nearly 
ideal hydrodynamic behavior~\cite{Ollitrault:1992bk,Heinz:2013th,Gale:2013da}. The second and third Fourier components of the particle azimuthal distribution, known as elliptic ($v_2$) and triangular ($v_3$) flow, respectively, 
reflect the response of the medium to the initial collision geometry, including its fluctuations, providing insights into the fundamental 
transport properties of the QGP medium~\cite{Schenke:2010rr,Qiu:2011hf}.  
Similar correlations have been observed in high-multiplicity proton-proton (\pp)~\cite{Khachatryan:2010gv,Aad:2015gqa,Khachatryan:2015lva,CMS:2016fnw} and  proton-lead (\pPb)~\cite{CMS:2013jlh,CMS:2012qk,CMS:2015yux,alice:2012qe,Aad:2012gla,Aaij:2015qcq} collisions at the LHC, 
as well as in deuteron-gold~\cite{Adare:2014keg,Adamczyk:2015xjc} and helium-3-gold~\cite{Adare:2015ctn} collisions at RHIC. The properties of the observed long-range correlations in these systems have been found to be consistent with hydrodynamic descriptions of a QGP droplet~\cite{Nagle:2018nvi,Dusling:2015gta}. 

For a transverse momentum (\pt) greater than $\sim$3\GeV, partons are not expected to be thermalized in the hot and dense medium created in heavy ion collisions, 
and a hydrodynamic picture is not applicable~\cite{Gale:2013da}. The observed azimuthal anisotropy of such particles in noncentral AA  collisions is primarily attributed to the energy loss of high-\pt partons as they traverse the QGP medium, whose geometry reflects the lenticular shape of the collision overlap region~\cite{Rudolph,Marquet:2009ae}. The extent of this energy loss depends on the path length, which, in turn, depends on the direction of the parton relative to the orientation of the medium~\cite{PhysRevLett.109.022301,Sirunyan:2017pan,proceedings2019010042,Filimonov:2002xk}. Therefore, the study of azimuthal anisotropy in small systems, where the overlap geometry is not expected to have a dominant role, is of considerable interest~\cite{CMS:2015yux,CMS:2015xmx}.

One of the challenges in analyzing azimuthal correlations, particularly in small colliding systems, is separating genuine collective effects from those arising from particles that are correlated as a result of other unrelated processes (such as jets and resonance decays). These so-called ``nonflow'' correlations that can be difficult to separate from the global flow anisotropies. The azimuthal correlations of high-\pt particles in small systems have been previously studied using two-particle correlations ~\cite{ATLAS:2019vcm,ALICE:2022cwa}. In those analyses, corrections for nonflow correlations are made using the relative multiplicity dependence of flow and nonflow correlations. The nonflow correction is then based on assumptions about the evolution of nonflow from low to high multiplicity collisions, and about the dominance of nonflow in low multiplicity collisions.

This Letter presents a measurement of elliptic anisotropies for high-\pt particles in \pPb collisions at a center-of-mass energy per nucleon pair $\sqrtsNN = 8.16\TeV$, compared to those in lead-lead (\PbPb) collisions at $\sqrtsNN  =  5.02\TeV$. The \pPb and \PbPb data were obtained using the CMS detector at the LHC in 2016 and 2018, corresponding to integrated luminosities of 186\nbinv and 0.607\nbinv, respectively~\cite{CMS:LUM-17-002,CMS:LUM-18-001}. The Q-cumulant multiparticle correlation technique is used based on ``subevents'' dividing particles into multiple pseudorapidity regions~\cite{Bilandzic:2010jr,DiFrancesco:2016srj,Jia:2017hbm}. By increasing the number of subevents beyond two, this approach helps in suppressing nonflow effects in a data-driven way. The results are presented as a function of \pt based on 2-, 3-, and 4-subevent analyses. Tabulated results for this analysis can be accessed in the HEPdata record~\cite{hepdata.156174}.

The central feature of the CMS apparatus~\cite{CMS:2008xjf,CMS:2023gfb} is a superconducting solenoid with an internal diameter of 6\unit{m}, providing a magnetic field of 3.8\unit{T}. Within the solenoid volume are a silicon pixel and strip tracker~\cite{Chatrchyan:2014fea,CMS:2021ime}, a lead tungstate crystal electromagnetic calorimeter~\cite{CMS-DP-2020-021}, and a brass and scintillator hadron calorimeter, each composed of a barrel and two endcap sections. For the run period relevant to this analysis, the silicon pixel and strip tracker provided coverage for $\abs{\eta} < 2.4$. The hadron forward (HF) calorimeter~\cite{Damgov:687382} extends the pseudorapidity coverage provided by the barrel and endcap detectors, with the two halves, one on each end, providing coverage in the range $2.9 < \abs{\eta} < 5.2$. The HF calorimeters are azimuthally subdivided into $20^{\circ}$ modular wedges, which are further segmented to form $0.175{\times}0.175$ $(\Delta\eta{\times}\Delta\phi)$ towers. The CMS apparatus can be used to detect and identify electrons, muons, photons, and hadrons~\cite{CMS:2020uim,CMS:2018rym,Chatrchyan:2014fea}. A more detailed description of the CMS detector can be found in Ref.~\cite{CMS:2008xjf}.

For \pPb collisions, the beam energies for the proton and lead nuclei were 6.5\TeV and 2.56\TeV per nucleon, respectively, 
resulting in a center-of-mass energy of $\sqrtsNN = 8.16\TeV$. Because of the asymmetric beam energies, the nucleon-nucleon center-of-mass
in the pPb collisions was not at rest with respect to the laboratory frame but was moving in the direction of the proton beam with a rapidity shift of 0.465.
Data were taken with the directions of the beams reversed for assessing possible systematic effects. The results in each direction were consistent with each other and were merged; results are reported using the convention that the positive pseudorapidity is defined by the proton-going direction.

To select high-multiplicity \pPb collisions, 
a dedicated high-multiplicity trigger was implemented using the CMS level-1 (L1)~\cite{Sirunyan:2020zal} 
and high-level trigger (HLT) systems~\cite{Khachatryan:2016bia}. The HLT uses fast versions of the offline software. To trigger minimum bias (MB) \pPb events, 
at least one of the two HF calorimeters was required to register an energy deposit above a threshold of 1\GeV. 
Additionally, at least one track with $\pt > 0.4\GeV$ in the pixel tracker at HLT level was required. 

In this analysis, \pPb events were required to have a primary vertex within 15\unit{cm} of the nominal interaction point along the beam axis and 0.2\unit{cm} 
in the transverse direction. 
At least two reconstructed tracks must be associated with the primary vertex.
Results are reported in ranges of \noff, corresponding to the measured track multiplicity with $\pt > 0.4\GeV$ and $\abs{\eta}<2.4$, before correcting for the tracking efficiency. 
Table~\ref{tab:nch_ntrkoffline} of \appname \ref{app:mapping} reports the mean multiplicity $\langle \noff \rangle$ and  multiplicity after correcting for tracking inefficiencies $\langle \ncorr \rangle$, for various \noff ranges. A trigger efficiency plateau above 95\% is observed in the $185 \leq \noff < 250$ range.

The \PbPb MB events are selected based on signals surpassing readout threshold of 3\GeV on both sides of the HF calorimeters. The \PbPb events are also required to have a primary vertex within 15\unit{cm} of the nominal interaction point along the beam axis 
and 0.2\unit{cm} in the transverse direction. 
Additionally, each HF detector must have at least two towers with an energy deposit of 4\GeV or more.
The trigger, event reconstruction, and track selections considered in this analysis are the same as in previous CMS 
correlation analyses~\cite{CMS:2013jlh, CMS:2015yux, CMS:2016fnw}. 
To perform correlation measurements, only tracks that originate from the primary vertex and meet the high-purity criteria discussed in Ref.~\cite{Chatrchyan:2014fea} 
are used. The trigger and event selection efficiencies are found to be about 98\% and above 95\% for \pPb and \PbPb, respectively. 
Furthermore, for both cases, the impact parameter significance, \ie, the parameter value divided by its uncertainty, of the tracks with respect to the primary vertex in the longitudinal and transverse directions 
must be less than 3, 
and the relative \pt uncertainty must be less than 10\% for the \pt range used.

To ensure high tracking efficiency and minimal background contamination, only tracks with $\abs{\eta} < 2.4$ and with $\pt > 0.3 \GeV$ for \pPb collisions
and $\pt > 0.5 \GeV$ for \PbPb collisions are used. 
The selected tracks are corrected for tracking efficiency and detector acceptance effects using simulated Monte Carlo 
samples generated by \HIJING v1.35~\cite{Gyulassy:1994ew} 
and \HYDJET 1.9~\cite{Lokhtin:2005px} for \pPb and \PbPb collisions, respectively. 
The response of the CMS detector to these simulated events is based on \GEANTfour~\cite{Agostinelli:2002hh}. 
In the measurement of correlations in \pPb collisions, multiple interactions (pileup) in the recorded event are a source of background. 
The average number of collisions per bunch crossing in the \pPb data varied between 0.1 and 0.25.\
The procedure used for identifying and rejecting pileup is similar to that described in Ref.~\cite{CMS:2013jlh}. The pileup is negligible in \PbPb collisions.

The $Q$-cumulant method, as encoded in the \textsc{mcorrelations} package~\cite{Bilandzic:2013kga}, is used to determine the $v_2$ value. 
This method has been used to measure  four-, six-, and eight-particle correlations
in previous CMS publications~\cite{CMS:2019wiy,Sirunyan:2017igb}, with the multiparticle correlations being inherently less sensitive to 
few-particle correlations,
such as those arising from jet fragmentation and back-to-back dijet correlations. In this method, the two- and four-particle 
correlators, $\langle \langle 2 \rangle \rangle$ and $\langle \langle 4 \rangle  \rangle$, 
are defined as
\ifthenelse{\boolean{cms@external}}
{
\begin{multline}
	\llangle {2} \rrangle = \llangle \re^{in(\phi_{1}-\phi_{2})} \rrangle \quad \text{and} \\ \quad
  \llangle {4} \rrangle = \llangle \re^{in(\phi_{1}+\phi_{2}-\phi_{3}-\phi_{4})} \rrangle.
  \label{eqn:corr_2p_4p}
\end{multline}
}
{
  \begin{equation}
        \llangle {2} \rrangle = \llangle \re^{in(\phi_{1}-\phi_{2})} \rrangle \quad \text{and} \quad
  \llangle {4} \rrangle = \llangle \re^{in(\phi_{1}+\phi_{2}-\phi_{3}-\phi_{4})} \rrangle.
  \label{eqn:corr_2p_4p}
\end{equation}
  }
Here, $\phi_{\mathrm{j}}$ (j = 1, \ldots, 4) represents the azimuthal angles of four 
distinct particles in an event, and the notation $\llangle\cdots\rrangle$ signifies that the average is 
calculated first over all particles in an event, then across all events. The integer $n$ denotes the order of Fourier harmonic, with $n=2$ corresponding to elliptic flow~\cite{Schenke:2010rr}. The four-particle cumulant $c_n\{4\}$ is given as
\begin{equation}
  c_n\{4\} = \llangle {4} \rrangle - 2 \langle \langle 2 \rangle \rangle^2.
  \label{eqn:cn_2p_4p}
\end{equation}

The multiparticle cumulant method correlates each particle of interest (POI) with reference flow particles (RFPs).
In this analysis, the RFPs for the cumulant method are charged hadrons within $\abs{\eta}<2.4$ and with $0.3 \ (0.5) < \pt < 3.0\GeV$ for \pPb (\PbPb) collisions. In order to reduce the contribution of minijets, an upper limit of $\pt < 3.0\GeV$ is used for the RFPs ~\cite{Wang:1992db}. The \pt ranges used for the POIs are given in Table~\ref{tab:mean_pt} of \appname \ref{app:meanpt}.

In the standard $Q$-cumulant analysis, the POI and RFP ranges overlap in $\eta$, potentially introducing nonflow contributions from short-range correlations. To suppress these effects, a subevent method~\cite{DiFrancesco:2016srj} is used, where particles are selected from different $\eta$ ranges, ensuring a pseudorapidity gap (difference in pseudorapidity values, $\Delta \eta$) between correlated particles. Statistical uncertainties are estimated using a bootstrap method~\cite{Efron:1979bxm}.
\ifthenelse{\boolean{cms@external}}
{
\begin{figure}[hbtp]
\centering
   \includegraphics[width=0.49\textwidth]{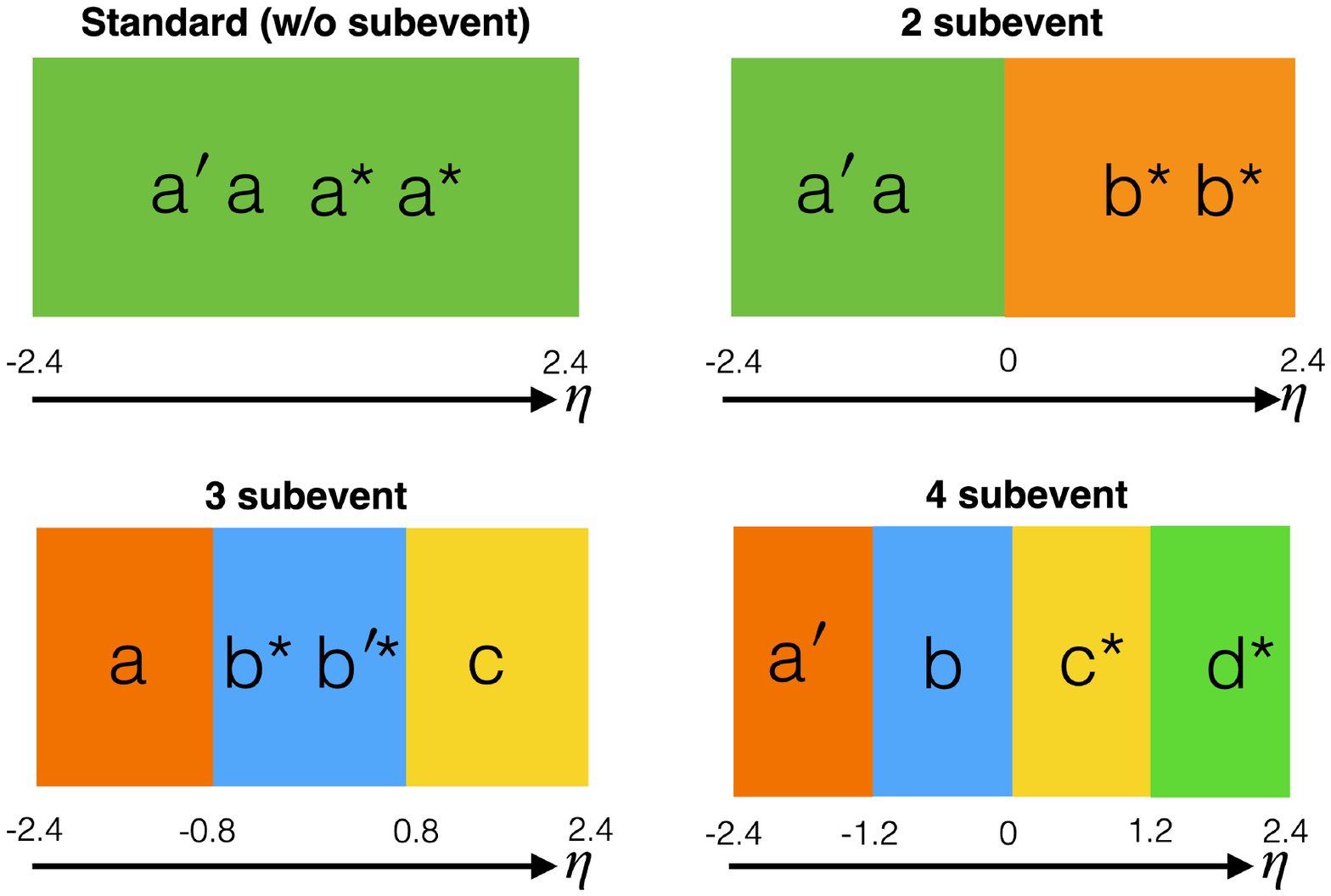}
   \caption{Different subevents in the CMS tracker acceptance used to correlate four particles. The denoted pseudorapidity ranges are used for the analysis. The letters denote the particles used in four-particle correlator. The asterisk (*) indicates particles with a negative azimuthal angle, and the prime symbol denotes the POI. 
   }
   \label{fig:subevent_pic}
\end{figure}
}
{
\begin{figure}[hbtp]
\centering
   \includegraphics[width=0.75\textwidth]{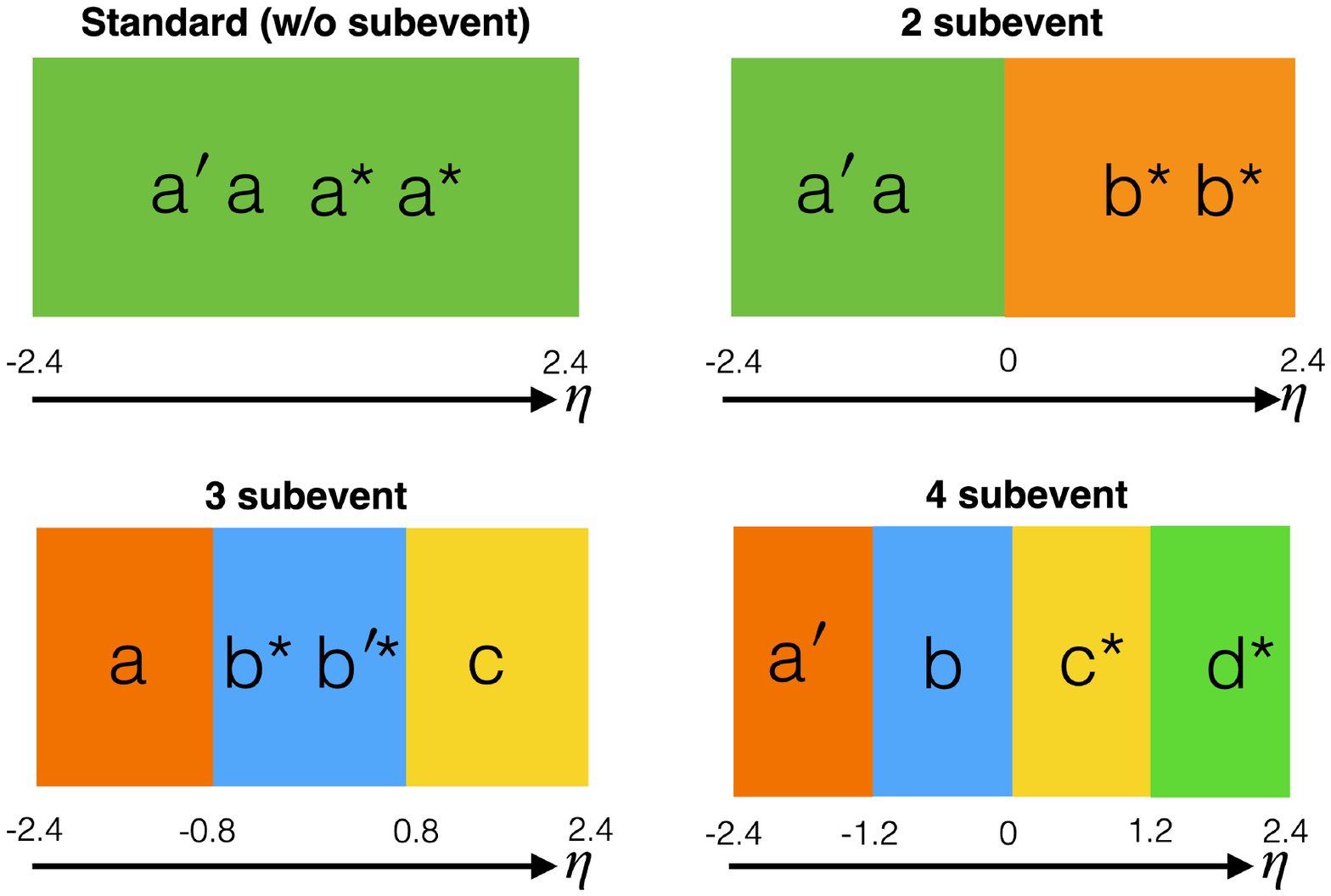}
   \caption{Different subevents in the CMS tracker acceptance used to correlate four particles. The denoted pseudorapidity ranges are used for the analysis. The letters denote the particles used in four-particle correlator. The asterisk (*) indicates particles with a negative azimuthal angle, and the prime symbol denotes the POI.
   }
   \label{fig:subevent_pic}
\end{figure}
}
Figure~\ref{fig:subevent_pic} illustrates how particle tracks are selected to construct four-particle cumulants. The upper left panel shows the case where all four particles are chosen from the same $\eta$ range, leading to significant nonflow contributions from single jets. The upper right panel depicts the two-subevent method ($-2.4<\eta<0$ and $0<\eta<2.4$), which reduces nonflow effects by separating particles $a$ and $b$ into distinct $\eta$ ranges. The three-subevent method (lower left) further suppresses dijet contributions by dividing into three $\eta$ ranges ($-2.4<\eta<-0.8$, $-0.8<\eta<0.8$ and $0.8<\eta<2.4$), while the four-subevent method (lower right) minimizes nonflow effects by assigning each particle to a distinct $\eta$ range ($-2.4<\eta<-1.2$, $-1.2<\eta<0$, $0<\eta<1.2$ and $1.2<\eta<2.4$) .

The \pt dependence of the flow coefficients can be studied using \pt-differential cumulants, $d_{n}\{4\}$, given by 
\begin{equation}
  \begin{aligned}
  d_{n}\{4\}=\langle \langle 4' \rangle^{} \rangle- 2\langle \langle 2' \rangle^{} \rangle  \langle \langle 2 \rangle^{} \rangle ,
  \label{dn4}
  \end{aligned}
\end{equation}
where $\langle \langle 4' \rangle^{} \rangle$ is a four-particle correlator with three RFPs and one POI and $\langle \langle 2' \rangle^{} \rangle$ 
is a two-particle correlator with one RFP and one POI.
In the following, we provide the specific $d_n\{4\}$ formulas for the two, three, and four subevent analyses. We use a notation where $a$, $b$, $c$, or $d$ 
refers to the particle chosen in a specific subevent and the symbol ``$\mid$'' indicates a separation between subevents. 
In Eq.(~\ref{2sub}), e.g., the notation $a'a\mid b^{*}b^{*}$ in the four-particle correlator means the first two particles are required to be in the first subevent ($a'a$) while the last two are required to be in the second subevent ($b^{*}b^{*}$), where $a'$ is the POI. Similarly, for the two-particle correlator, one particle in each subevent is required ($a'\mid b$).

The two-subevent differential cumulant can be expressed in terms of the correlators, with
\begin{equation}
  \begin{aligned}
 d_{n}\{4\}_{\rm{2sub}}=\langle \langle 4 \rangle^{a'a|b^{*}b^{*}} \rangle- 2\langle \langle 2 \rangle^{a'|b^{*}} \rangle  \langle \langle 2 \rangle^{a|b^{*}} \rangle  .
\label{2sub}
\end{aligned}
\end{equation}
For the three subevent case, the correlator expansion depends on the POI pseudorapidity range, with
\begin{equation}
  \begin{aligned}
  d_{n}\{4\}_{\rm{3sub}} \ =\langle \langle 4 \rangle^{a|b^{*}b^{*}|c'} \rangle- 2\langle \langle 2 \rangle^{a|b^{*}} \rangle  \langle \langle 2 \rangle^{b^{*}|c'} \rangle ,
\label{3subfb}
\end{aligned}
\end{equation}
when the POI is in the range $0.8< \abs{\eta} <2.4$ and
\ifthenelse{\boolean{cms@external}}
{
\begin{multline}
  \begin{aligned}
d_{n}\{4\}_{\rm{3sub}} \ =\langle \langle 4 \rangle^{a|b^{*}b'^{*}|c} \rangle- \langle \langle 2 \rangle^{a|b'^{*}} \rangle  \langle \langle 2 \rangle^{b^{*}|c} \rangle- \\ \langle \langle 2 \rangle^{a|b^{*}} \rangle  \langle \langle 2 \rangle^{b'^{*}|c} \rangle ,
\label{3subm}
\end{aligned}
\end{multline}
}
{
\begin{equation}
  \begin{aligned}
d_{n}\{4\}_{\rm{3sub}} \ =\langle \langle 4 \rangle^{a|b^{*}b'^{*}|c} \rangle- \langle \langle 2 \rangle^{a|b'^{*}} \rangle  \langle \langle 2 \rangle^{b^{*}|c} \rangle- \langle \langle 2 \rangle^{a|b^{*}} \rangle  \langle \langle 2 \rangle^{b'^{*}|c} \rangle ,
\label{3subm}
\end{aligned}
\end{equation}
}  
when $\abs{\eta}<0.8$. 
In the latter case, we have two choices for the POI and, accordingly, two combinations for the product of two-particle correlators 
$(\langle \langle 2 \rangle^{a|b'^{*}} \rangle  \langle \langle 2 \rangle^{b^{*}|c} \rangle$ and 
$\langle \langle 2 \rangle^{a|b^{*}} \rangle  \langle \langle 2 \rangle^{b'^{*}|c} \rangle)$. 
Similarly, in the case of four subevents, instead of taking twice the product of two-particle correlators, we incorporate two different choices 
($\langle \langle 2 \rangle^{a'|c^{*}} \rangle  \langle \langle 2 \rangle^{b|d^{*}} \rangle$ and 
$\langle \langle 2 \rangle^{a'|d^{*}} \rangle  \langle \langle 2 \rangle^{b|c^{*}} \rangle$~), with
\ifthenelse{\boolean{cms@external}}
{
\begin{multline}
  \begin{aligned}
d_{n}\{4\}_{\rm{4sub}} \ =\langle \langle 4 \rangle^{a'|b|c^{*}|d^{*}} \rangle- \langle \langle 2 \rangle^{a'|c^{*}} \rangle  \langle \langle 2 \rangle^{b|d^{*}} \rangle- \\ \langle \langle 2 \rangle^{a'|d^{*}} \rangle  \langle \langle 2 \rangle^{b|c^{*}} \rangle 
\label{4sub}
\end{aligned}
\end{multline}
}
{
\begin{equation}
  \begin{aligned}
d_{n}\{4\}_{\rm{4sub}} \ =\langle \langle 4 \rangle^{a'|b|c^{*}|d^{*}} \rangle- \langle \langle 2 \rangle^{a'|c^{*}} \rangle  \langle \langle 2 \rangle^{b|d^{*}} \rangle-\langle \langle 2 \rangle^{a'|d^{*}} \rangle  \langle \langle 2 \rangle^{b|c^{*}} \rangle
\label{4sub}
\end{aligned}
\end{equation}
}
when the POI is in the range $-2.4<\eta<-1.2$ and similar equations for when the POI is in different $\eta$ ranges.
The differential $v_{n}\{4\}$ value can then be expressed as
\begin{equation}
  \begin{aligned}
  v_{n}\{4\}= \ - \ d_{n}\{4\}/(-c_{n}\{4\})^{3/4},
  \label{vn4}
\end{aligned}
\end{equation}
where $c_n\{4\}$ is the integral cumulant for RFPs~\cite{Bilandzic:2010jr}. The effectiveness of four-subevent method in nonflow removal is confirmed by a study done using generator level \HIJING events (Figure~\ref{fig:d24_4sub_hij} of \appname \ref{app:hijing}), where collective effects are absent and we expect zero values after nonflow removal.

The systematic uncertainties are evaluated as functions of \pt using alternative procedures for extracting
the cumulants, following the Barlow procedure~\cite{Barlow:2002yb}. Systematic uncertainties due to tracking inefficiency and track misreconstruction rate are studied by varying the track quality requirements. 
The selection thresholds on
the significances of the transverse and longitudinal track impact parameters 
are varied from 2 to 5. In addition, the upper limit on the relative \pt uncertainty
is varied from 5 to 10\% (15\%) for \pPb (\PbPb) collisions. The resulting systematic uncertainty
varies up to 5\% depending on multiplicity, \pt, and collision system. The potential systematic uncertainty arising from the dependence of the acceptance and reconstruction efficiency on the primary vertex position along the beam axis ($z_\text{vtx}$) is determined by comparing the results with events from $\abs{z_\text{vtx}}<3$\unit{cm} and $3<\abs{z_\text{vtx}}<15$\unit{cm} to the default range of $\abs{z_\text{vtx}}<15$\unit{cm}. This systematic uncertainty is estimated to be $<$0.2\%. The impact of pileup in \pPb collisions is studied by varying the pileup selection of events using the distance among reconstructed vertices and their associated number of tracks. 
The corresponding systematic uncertainty is also found to be $<0.1\%$. Systematic effects from event selections in \PbPb collisions are found to be negligible. 
Systematic uncertainty from the different sources are added in quadrature to obtain the total systematic uncertainty 
shown as boxes in the figures.

\ifthenelse{\boolean{cms@external}}
{
\begin{figure}[hbt]
\centering
    \includegraphics[width=0.49\textwidth]{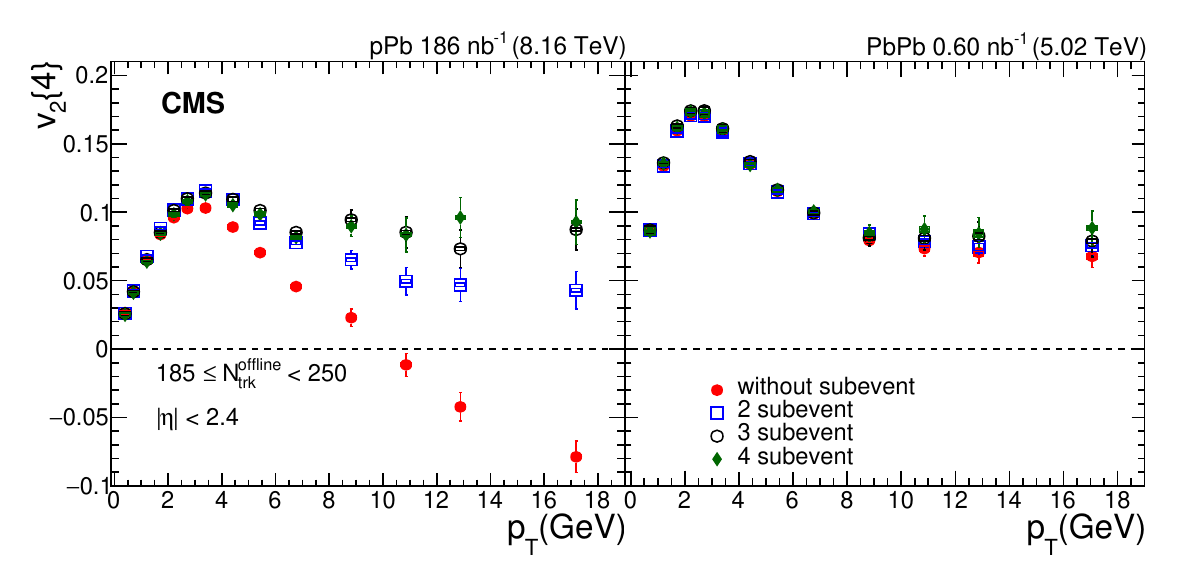}
  \caption{ The $v_{2}\{4\}$ value \vs \pt with $185 \leq \noff < 250$ for \pPb (left) and \PbPb (right) collisions. The red, blue, black and green markers are for the cases where tracks are
not divided into subevents, and divided into two, three and four subevents,
respectively. The solid lines and boxes represent the statistical and systematic uncertainties, respectively.  }
  \label{fig:v24_nsub_4sub}
\end{figure}
}
{
\begin{figure}[!h]
\centering
    \includegraphics[width=0.96\textwidth]{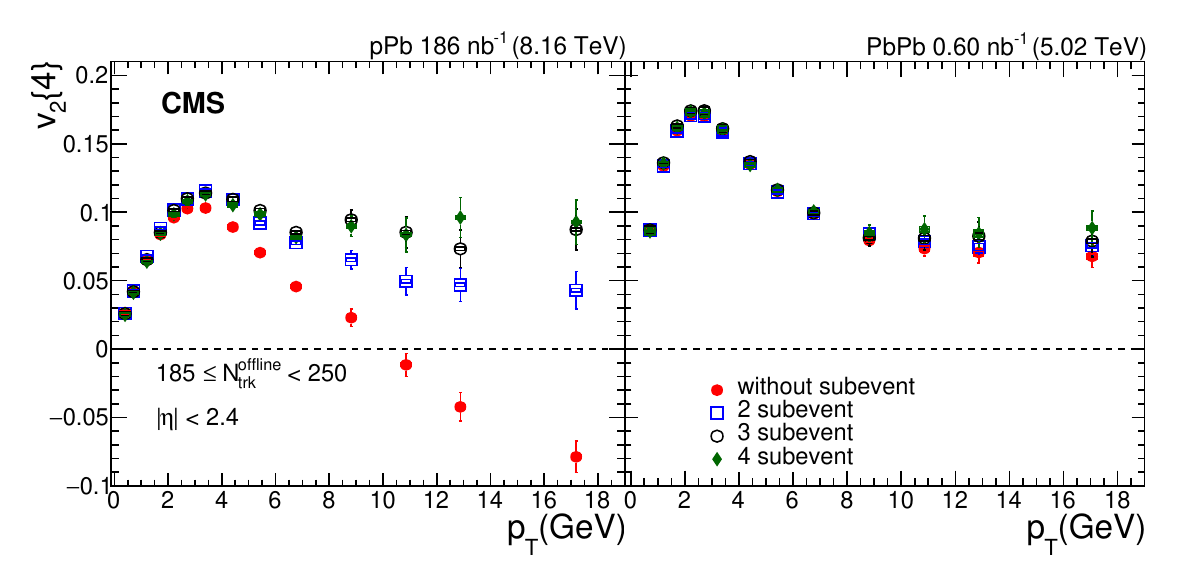}
  \caption{ The $v_{2}\{4\}$ value \vs \pt with $185 \leq \noff < 250$ for \pPb (left) and \PbPb (right) collisions. The red, blue, black and green markers are for the cases where tracks are
not divided into subevents, and divided into two, three and four subevents,
respectively. The solid lines and boxes represent the statistical and systematic uncertainties, respectively.  }
  \label{fig:v24_nsub_4sub}
\end{figure}
}
\begin{figure}[hbt]
\centering
    
    \includegraphics[width=0.49\textwidth]{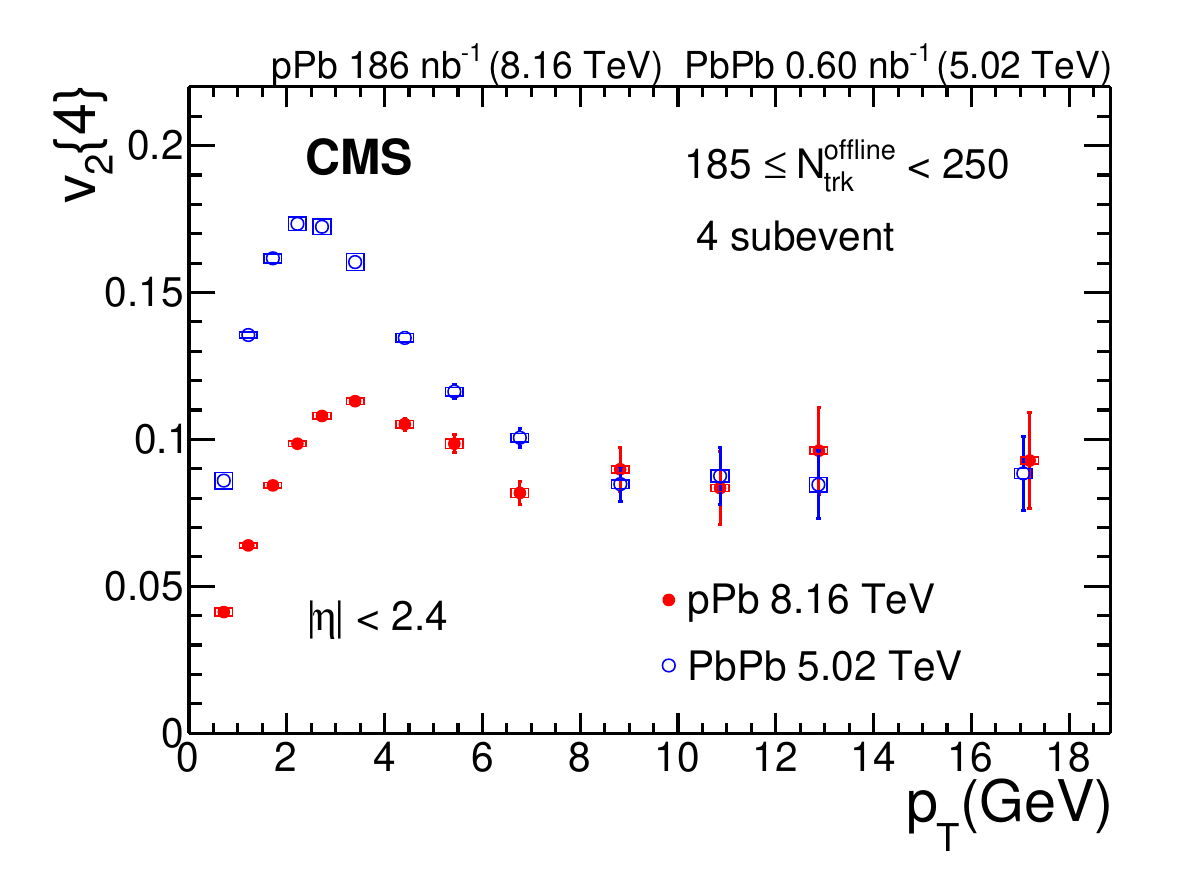}
    \includegraphics[width=0.49\textwidth]{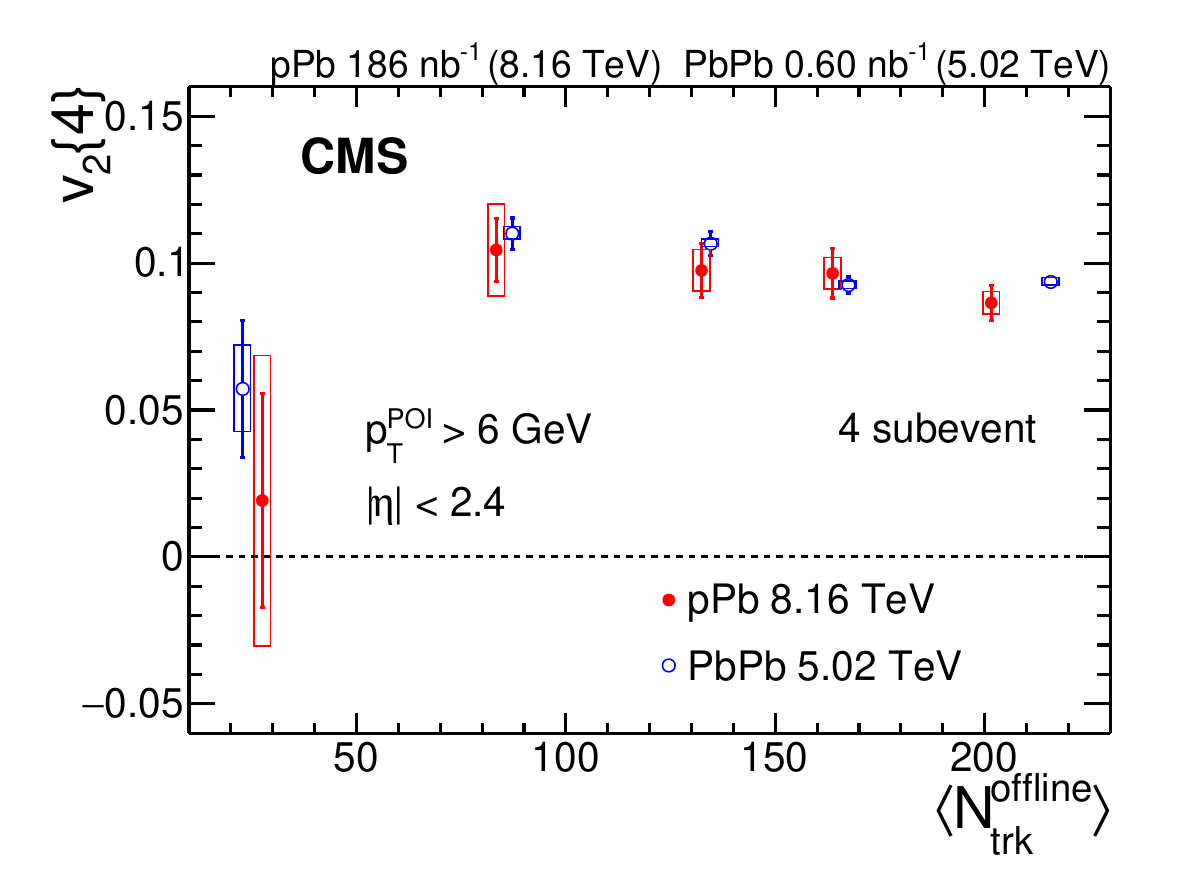}
  \caption{\cmsLeft:  Comparison of \pPb and \PbPb four-subevent $v_{2}\{4\}$ values with $185 \leq \noff < 250$ \vs \pt. 
	   \cmsRight: Comparison of \pPb and \PbPb four-subevent $v_{2}\{4\}$ values 
	with $\ptpoi > 6\GeV$ \vs $\langle \noff \rangle$. 
	The solid lines and boxes represent the statistical and systematic uncertainties, respectively.  }
  \label{fig:v24_4sub_a6}
\end{figure}

The $v_{2}\{4\}$ results for the different subevent methods are shown for the $185 \leq \noff < 250$ range as function of \pt  
in Fig.~\ref{fig:v24_nsub_4sub}. The left panel shows the results for \pPb collisions and the right panel for \PbPb collisions. The $185 \leq \noff < 250$ range corresponds to approximately 65--70\% centrality in \PbPb collisions. The $v_{2}\{4\}$ is calculated for each \pt range of POI and the marker is placed at the mean \pt value.
The standard $v_{2}\{4\}$ result (without subevents) becomes negative above $\pt \sim 10\GeV$ for \pPb collisions, 
which is not the case for the \PbPb results. 
This is likely a consequence of selecting rare high-multiplicity \pPb events with an increased nonflow 
contribution from jets, which results in a negative contribution to $v_{2}\{4\}$, when compared with \PbPb events. 
Using the subevent method, this negative trend is heavily suppressed and positive values of $v_{2}\{4\}$ 
(negative value of $d_{2}\{4\}$) result up to $\pt \sim 17$\GeV, the upper limit of the current investigation. 
The three- and four-subevent results are in agreement with each other, suggesting a nearly complete removal of short range correlations.

The \cmsLeft plot in Fig.~\ref{fig:v24_4sub_a6} compares the four-subevent $v_{2}\{4\}$ values as a function of \pt for \pPb and \PbPb collisions. 
We observe similar magnitudes of $v_{2}\{4\}$ in both collision systems at higher \pt, 
with nearly constant values above $\pt \sim 6$\GeV. In the \cmsRight plot of Fig.~\ref{fig:v24_4sub_a6}, 
we show the four-subevent $v_{2}\{4\}$ for particles with $\pt > 6\GeV$ as a function of $\langle \noff \rangle$ for \pPb and \PbPb collisions. 
Figure~\ref{fig:v24_4sub_a6_corr} of \appname \ref{app:expfigure} shows the $v_{2}\{4\}$ results in terms of efficiency-corrected mean multiplicity ($\langle \ncorr \rangle$). 
The $v_{2}\{4\}$ values and their $\langle \noff \rangle$ dependence for the two systems are again very similar. This similarity is unlikely to be a consequence of the different \sqrtsNN values of the two systems being considered, noting that only minimal (2 to 4\%) difference in integrated $v_{2}$ values is observed in \PbPb collisions at $\sqrtsNN  =  2.76\TeV$ and $\sqrtsNN  =  5.02\TeV$~\cite{ALICE:2018rtz}.

The four-subevent $v_{2}\{4\}$ values at high \pt are significantly larger than the previous measurement of $v_{2}\{2\}$ by the ATLAS experiment using the template fit method~\cite{ATLAS:2019vcm}. Using the same template fit method, we obtain consistent $v_{2}\{2\}$ results compared to ATLAS. The template fit method assumes that the shape of nonflow (such as jet-induced) correlations remain unchanged between low and high multiplicity events. This assumption may, however, not hold because of angular broadening of these correlations with increasing multiplicity, potentially leading to an over-subtraction of nonflow correlations in the genuine $v_{2}$ signal. We have confirmed this by analyzing $v_{2}\{2\}$ results in \HIJING simulations using the template fit method.

Our results are unexpected, given the distinct nature and geometry of the medium formed in \pPb as compared to \PbPb collisions. 
The baryon-meson ratio enhancement found at intermediate \pt disappears for $\pt > 6\GeV$ in \PbPb collisions~\cite{ALICE:2013cdo}. This is consistent with the higher-\pt region studied here being well within the hard scattering regime where 
any azimuthal asymmetry is expected to arise through an energy loss mechanism, as opposed to hydrodynamic flow. High-\pt hadrons are significantly suppressed in \PbPb collisions relative to \pp collisions, 
indicating a path length dependent energy loss~\cite{CMS:2016xef}. No such energy loss is apparent in \pPb collisions, 
although it is important to note that nuclear modification measurements in small systems are complicated by uncertainties in the calculated number of binary nucleon-nucleon collisions~\cite{PHENIX:2013jxf,STAR:2014qsy}. 
The larger $v_{2}\{4\}$ values observed for low multiplicity \PbPb collisions below $\pt \sim 3$\GeV suggest a more lenticular collision geometry 
when compared to high-multiplicity \pPb collisions. This, together with the energy loss argument, leads to an expectation of larger  $v_{2}\{4\}$ values at high \pt for \PbPb events.  
However, our observed high-\pt $v_{2}\{4\}$ values are similar for \PbPb and \pPb collisions. One potential origin of the $v_{2}\{4\}$ at high \pt could be the initial-state intrinsic transverse momentum asymmetries of unpolarized and polarized partons within an unpolarized nucleon, as explored in Ref.~\cite{Soudi:2023epi}. Further investigations, both in theory and in experiment are needed.

In summary, four-particle elliptic anisotropies ($v_2\{4\}$) are measured over a large range of transverse momentum (\pt) using multiparticle cumulant analyses 
based on the division of particle tracks into two, three, or four separate pseudorapidity ranges (``subevents''). Dividing the particles into separate subevents is shown to strongly suppress effects resulting from jet fragmentation. Significant and positive  $v_2\{4\}$ values that extend to high \pt are found for high-multiplicity \pPb events. Results are shown  for proton-lead (\pPb) and lead-lead (\PbPb) collisions at nucleon-nucleon center-of-mass energies of 8.16\TeV and  5.02\TeV, respectively. Similar magnitudes of $v_{2}\{4\}$ values at $\pt > 8 \GeV$ are observed in \pPb and \PbPb collisions at comparable multiplicities. The observed agreement in multi-particle collectivity for high-\pt particles in the two systems is striking and may hint at a similar origin.

\begin{acknowledgments}
We congratulate our colleagues in the CERN accelerator departments for the excellent performance of the LHC and thank the technical and administrative staffs at CERN and at other CMS institutes for their contributions to the success of the CMS effort. In addition, we gratefully acknowledge the computing centers and personnel of the Worldwide LHC Computing Grid and other centers for delivering so effectively the computing infrastructure essential to our analyses. Finally, we acknowledge the enduring support for the construction and operation of the LHC, the CMS detector, and the supporting computing infrastructure provided by the following funding agencies: SC (Armenia), BMBWF and FWF (Austria); FNRS and FWO (Belgium); CNPq, CAPES, FAPERJ, FAPERGS, and FAPESP (Brazil); MES and BNSF (Bulgaria); CERN; CAS, MoST, and NSFC (China); MINCIENCIAS (Colombia); MSES and CSF (Croatia); RIF (Cyprus); SENESCYT (Ecuador); ERC PRG, RVTT3 and MoER TK202 (Estonia); Academy of Finland, MEC, and HIP (Finland); CEA and CNRS/IN2P3 (France); SRNSF (Georgia); BMBF, DFG, and HGF (Germany); GSRI (Greece); NKFIH (Hungary); DAE and DST (India); IPM (Iran); SFI (Ireland); INFN (Italy); MSIP and NRF (Republic of Korea); MES (Latvia); LMTLT (Lithuania); MOE and UM (Malaysia); BUAP, CINVESTAV, CONACYT, LNS, SEP, and UASLP-FAI (Mexico); MOS (Montenegro); MBIE (New Zealand); PAEC (Pakistan); MES and NSC (Poland); FCT (Portugal); MESTD (Serbia); MCIN/AEI and PCTI (Spain); MOSTR (Sri Lanka); Swiss Funding Agencies (Switzerland); MST (Taipei); MHESI and NSTDA (Thailand); TUBITAK and TENMAK (Turkey); NASU (Ukraine); STFC (United Kingdom); DOE and NSF (USA).
\end{acknowledgments}

\bibliography{auto_generated} 
\clearpage
\appendix
\ifthenelse{\boolean{cms@external}}{\section*{Back matter}}{\section{Back matter}}

\subsection{Mapping between \texorpdfstring{\noff}{Ntrkoffline} and \texorpdfstring{\ncorr}{Ntrkcorrected} in \texorpdfstring{\pPb}{pPb} and \texorpdfstring{\PbPb}{PbPb} collisions\label{app:mapping}}
Table~\ref{tab:nch_ntrkoffline} gives Average multiplicity ($\langle \noff \rangle$) and multiplicity corrected for tracking efficiency ($\langle \ncorr \rangle$) values in various \noff ranges for \pPb and \PbPb collisions. Uncertainties for the tracking efficiency corrected \ncorr are included.
\renewcommand{\arraystretch}{1.2}
\begin{table}[ht]
\centering
\topcaption{\label{tab:nch_ntrkoffline} $\langle \noff \rangle$ and $\langle \ncorr \rangle$ values in various \noff ranges.}
\begin{scotch}{ c c c {c}@{\hspace*{5pt}} c   c }
\multicolumn{1}{l}{} & \multicolumn{2}{c}{pPb} &&                   \multicolumn{2}{c}{PbPb}\\
\cline{2-3}\cline{5-6}
\noff range & $\langle \noff \rangle$ & $\langle \ncorr \rangle$ & & $\langle \noff \rangle$ &                  $\langle \ncorr \rangle$ \\
\hline
$(0, 60)$      & 27    & 33$\pm$1 & & 23    & 39$\pm$2     \\
$[60, 120)$    & 83   & 101$\pm$4 & & 87    & 152$\pm$6      \\
$[120, 150)$   & 132  & 160$\pm$6  && 135    & 233$\pm$10      \\
$[150, 185)$   & 164   & 198$\pm$7  && 168    & 287$\pm$12      \\
$[185, 250)$   & 202  & 245$\pm$10  && 216    & 368$\pm$16     \\
\end{scotch}
\end{table}

\subsection{Mean \texorpdfstring{\pt}{pt} for different \texorpdfstring{\pt}{pt} bins in \texorpdfstring{\pPb}{pPb} and \texorpdfstring{\PbPb}{PbPb} for \texorpdfstring{$185 \leq \noff < 250$}{185 <= Ntrkoffline < 250}\label{app:meanpt}}
Table ~\ref{tab:mean_pt} gives the mean \pt ($\langle \pt \rangle$) values for various \pt ranges used for the POIs in \pPb and \PbPb collisions. The difference is negligible except the last bin.
\renewcommand{\arraystretch}{1.2}
\begin{table}[htb!]
\centering
\topcaption{\label{tab:mean_pt} $\langle \pt \rangle$ for \pt bins of POI in \pPb and \PbPb for $185 \leq \noff < 250$. All \pt values are given in units of \GeV.}
\begin{scotch}{ c c {c}@{\hspace*{5pt}} c }
 & pPb && PbPb\\
\cline{2-2}\cline{4-4}
\pt range & $\langle \pt \rangle$ && $\langle \pt \rangle$ \\
\hline
0.3-0.5     & 0.40   &&  \\
0.5-1.0     & 0.71 && 0.71 \\
1.0-1.5     & 1.21 && 1.21 \\
1.5-2.0     & 1.71 && 1.71 \\
2.0-2.5     & 2.22 && 2.22 \\
2.5-3.0     & 2.72 && 2.72 \\
3.0-4.0     & 3.40 && 3.40 \\
4.0-5.0     & 4.41 && 4.41 \\
5.0-6.0     & 5.43 && 5.42 \\
6.0-8.0     & 6.77 && 6.76 \\
8.0-10.0    & 8.81 && 8.81 \\
10.0-12.0   & 10.86 && 10.85 \\
12.0-14.0   & 12.88 && 12.87 \\
$>$14.0     & 17.19 && 17.06 \\
\end{scotch}
\end{table}

\subsection{Comparison of 4-subevent \texorpdfstring{$d_{2}\{4\}$}{d2{4}} in \texorpdfstring{\HIJING}{HIJING} with data }
\label{app:hijing}
\begin{figure}[htb!]
\centering
    \includegraphics[width=0.49\textwidth]{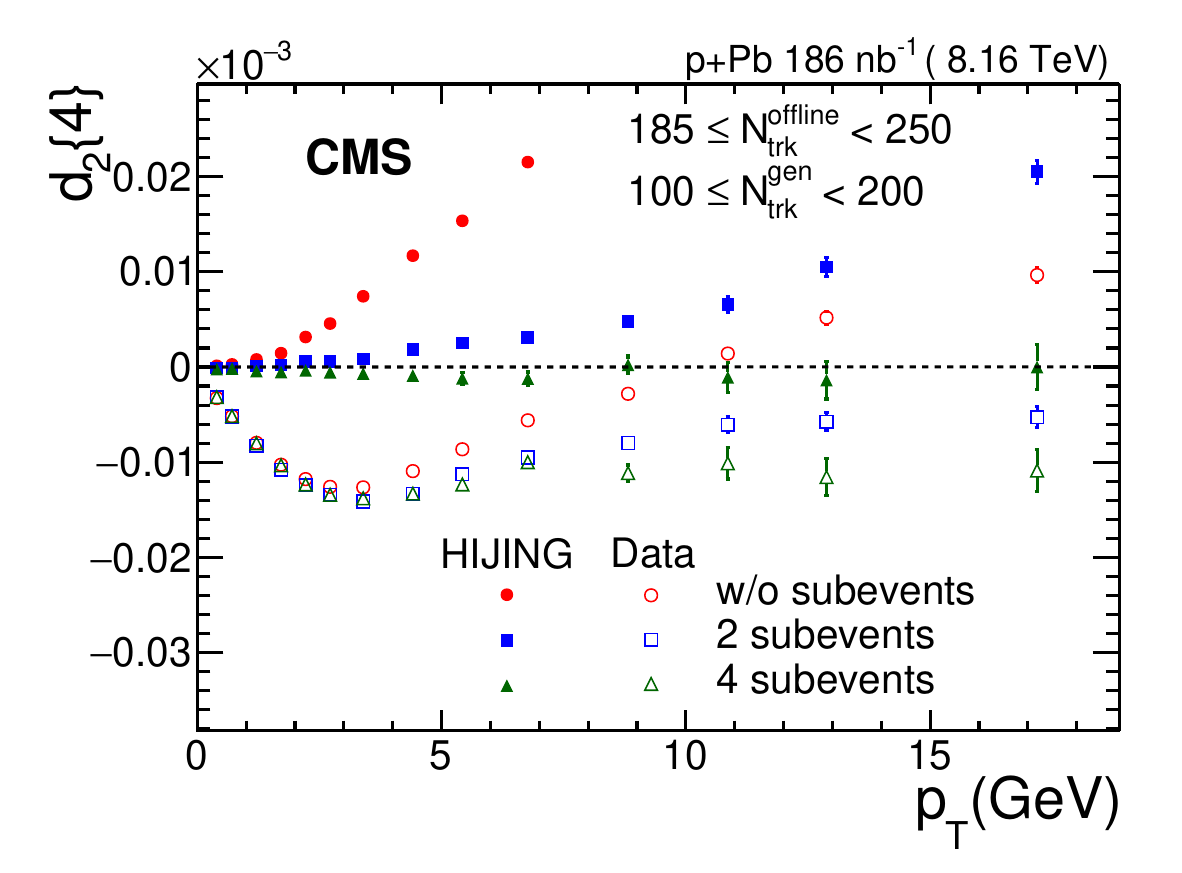}
  \caption{
        The four-subevent $d_{2}\{4\}$ values in generator level \HIJING for $60 \leq \ngen < 120$ shown by the closed green diamonds are compared to the experimental \pPb results with $60 \leq \noff < 120$ (open blue squares) and $185 \leq \noff < 250$ (closed red circles).
        }
  \label{fig:d24_4sub_hij}
\end{figure}
Figure~\ref{fig:d24_4sub_hij} compares $d_{2}\{4\}$ results based on \HIJING simulations in $100 \leq \ngen < 200$ with the experimental data results in $185 \leq \noff < 250$. Here \ngen is the measured track multiplicity with $\pt > 0.4\GeV$ and $\abs{\eta}<2.4$ in generator level \HIJING. The $d_{2}\{4\}$ decreases significantly as we go from ``without subevents'' to ``4 subevents'' due to reduction in nonflow in both HIJING (filled points) and data (open points). Although the amount of reduction is smaller in data than in HIJING, the systematic trends are the same. The \HIJING results of 4 subevent $d_{2}\{4\}$ are consistent with zero at high \pt, suggesting complete removal of nonflow. Whereas, the experimental data results of 4 subevent $d_{2}\{4\}$ are found to have significantly negative values.

\ifthenelse{\boolean{cms@external}}{\clearpage}{}
\subsection{Plot shown in Fig.~\ref{fig:v24_4sub_a6} (right), but with \texorpdfstring{$\langle \ncorr \rangle$}{<Ntrkcorrected>} along the x-axis. }
\label{app:expfigure}
\begin{figure}[htb!]
\centering
    \includegraphics[width=0.49\textwidth]{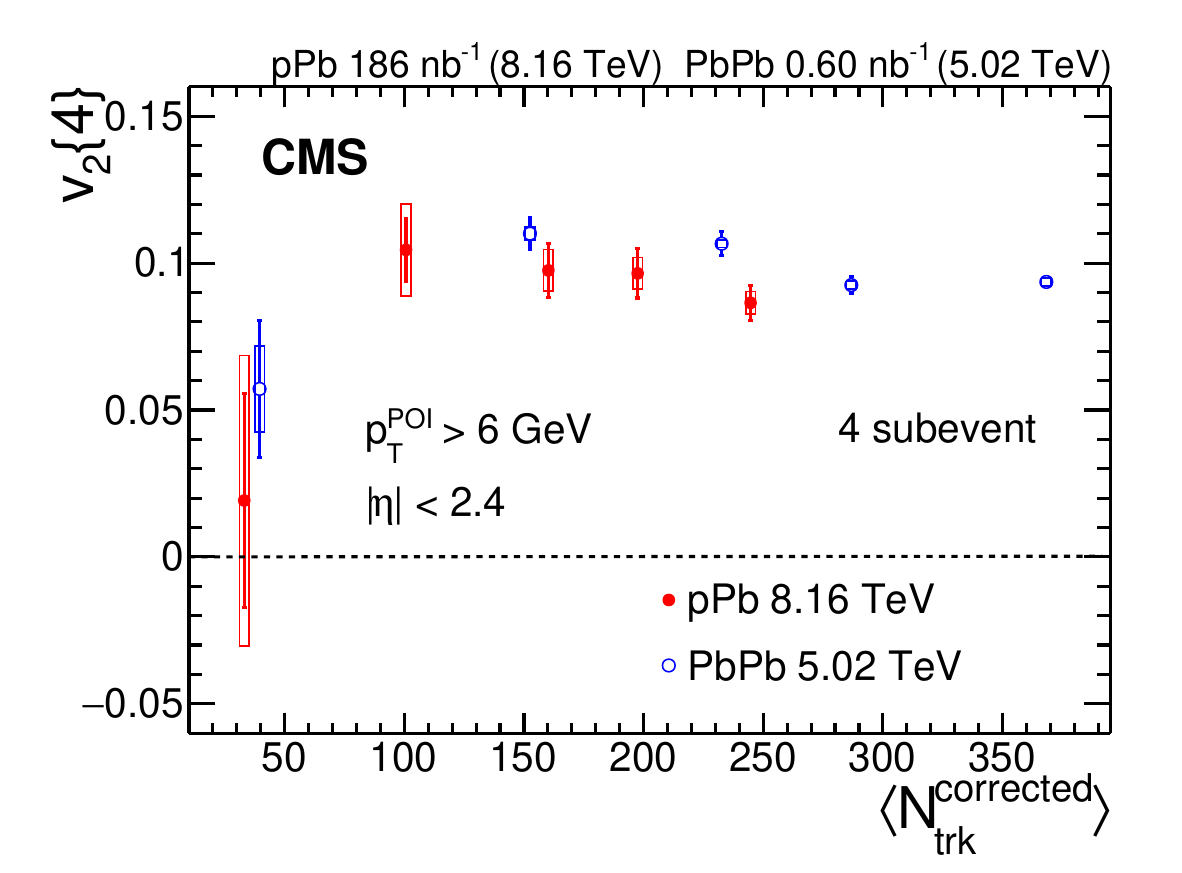}
  \caption{ 
	Comparison of $v_{2}\{4\}$ values as a function of $\langle \ncorr \rangle$ between \pPb and \PbPb collisions. The solid lines and boxes indicate the statistical and systematic uncertainties, respectively.
	}
  \label{fig:v24_4sub_a6_corr}
\end{figure}

Figure~\ref{fig:v24_4sub_a6_corr} shows the four-subevent $v_{2}\{4\}$ results in terms of efficiency-corrected mean multiplicity ($\langle \ncorr \rangle$) in \pPb and \PbPb collisions. The POI is taken with $\pt > 6 \GeV$. The high-\pt $v_{2}\{4\}$ values and their $\langle \noff \rangle$ dependence for the two systems are very similar.

\cleardoublepage \section{The CMS Collaboration \label{app:collab}}\begin{sloppypar}\hyphenpenalty=5000\widowpenalty=500\clubpenalty=5000
\cmsinstitute{Yerevan Physics Institute, Yerevan, Armenia}
{\tolerance=6000
V.~Chekhovsky, A.~Hayrapetyan, V.~Makarenko\cmsorcid{0000-0002-8406-8605}, A.~Tumasyan\cmsAuthorMark{1}\cmsorcid{0009-0000-0684-6742}
\par}
\cmsinstitute{Institut f\"{u}r Hochenergiephysik, Vienna, Austria}
{\tolerance=6000
W.~Adam\cmsorcid{0000-0001-9099-4341}, J.W.~Andrejkovic, L.~Benato\cmsorcid{0000-0001-5135-7489}, T.~Bergauer\cmsorcid{0000-0002-5786-0293}, S.~Chatterjee\cmsorcid{0000-0003-2660-0349}, K.~Damanakis\cmsorcid{0000-0001-5389-2872}, M.~Dragicevic\cmsorcid{0000-0003-1967-6783}, P.S.~Hussain\cmsorcid{0000-0002-4825-5278}, M.~Jeitler\cmsAuthorMark{2}\cmsorcid{0000-0002-5141-9560}, N.~Krammer\cmsorcid{0000-0002-0548-0985}, A.~Li\cmsorcid{0000-0002-4547-116X}, D.~Liko\cmsorcid{0000-0002-3380-473X}, I.~Mikulec\cmsorcid{0000-0003-0385-2746}, J.~Schieck\cmsAuthorMark{2}\cmsorcid{0000-0002-1058-8093}, R.~Sch\"{o}fbeck\cmsAuthorMark{2}\cmsorcid{0000-0002-2332-8784}, D.~Schwarz\cmsorcid{0000-0002-3821-7331}, M.~Sonawane\cmsorcid{0000-0003-0510-7010}, W.~Waltenberger\cmsorcid{0000-0002-6215-7228}, C.-E.~Wulz\cmsAuthorMark{2}\cmsorcid{0000-0001-9226-5812}
% ...
\par}
\vskip\cmsinstskip
\dag:~Deceased\\
$^{1}$Also at Yerevan State University, Yerevan, Armenia\\
$^{2}$Also at TU Wien, Vienna, Austria\\
% ...
\end{sloppypar}
%%% END EDITABLE REGION %%%
% skeleton_end
\end{document}